\documentstyle[12pt]{article}
\newcommand{\Z}{{\sf Z \!\!\! Z}}

\makeatletter
\@addtoreset{equation}{section}
\makeatother
\title{Capillary Waves in Binder's Approach to the Interface Tension}
\author{U.-J. Wiese$^+$ \\[2em]
Universit\"at Bern, Sidlerstrasse 5, 3012 Bern, Switzerland}
\begin{document}

\maketitle

\begin{abstract} \normalsize

In Binder's approach the reduced interface tension $\sigma$
of the Ising model in the broken phase
is determined from the finite volume effects of the partition function
$Z(M)$ at fixed total magnetization $M$. For small $|M|$ the partition function
of a system of size $L^d$ with periodic boundary conditions
is dominated by configurations with two interfaces, such that
$Z(M) \propto \exp(- 2 \sigma L^{d-1})$. Capillary wave fluctuations of the
interfaces correct this result to $Z(M) \propto L^x \exp(- 2 \sigma L^{d-1})$
with $x = -1$.
The knowledge of the pre-exponential behavior allows an improved fit of
numerical data, and a determination of the interface stiffness.

\end{abstract}

\vspace{5cm}
\begin{flushleft}
$^+$ supported by Schweizer Nationalfond
\end{flushleft}
\pagebreak

Let us consider the $d$-dimensional Ising model on a periodic $L^d$ lattice
in the low temperature broken phase.
Recent numerical simulations of the so-called multimagnetical ensemble
in the $d = 2$ and in the $d = 3$ case \cite{Ber91}
have demonstrated that Binder's approach \cite{Bin81} to the interface
tension is numerically feasible. The numerical method computes the partition
function as a function of the total magnetization
$M = \sum_x \sigma(x)$, where $\sigma(x) = \pm 1$ are the spin variables. The
partition function at fixed total magnetization is given by
\begin{equation}
Z(M) = \prod_x \sum_{\sigma(x) = \pm 1} \delta_{M,\sum_x \sigma(x)}
\exp(- \beta{\cal H}[\sigma]),
\end{equation}
where $\delta_{M,N}$ is the Kronecker $\delta$-function, $\beta$ is the
inverse temperature and ${\cal H}[\sigma] = - \sum_{\langle x y \rangle}
\sigma(x) \sigma(y)$ is the Hamilton function with nearest neighbor
interactions. When $|M|$ is small,
typical configurations consist of two domains (of $+$ and $-$ phase) separated
by two interfaces, which are closed via periodic boundary conditions.
The dynamics of the interfaces can be described in a capillary wave model.
In the $d = 2$ case the interfaces are lines, which one
can parametrize as $x_1(t)$ and $x_2(t)$ with $t \in [0,L]$. A typical
configuration is depicted in fig.1. \begin{figure}[h]
\vspace{11cm}
\caption{A typical configuration consisting of two domains of $+$ and $-$
phase separated by two interfaces parametrized as $x_1(t)$ and $x_2(t)$.}
\end{figure}

The energy of the configuration can be
written as
\begin{equation}
{\cal H}[x_1,x_2] = \int_0^L dt [fL +
\alpha(\varphi_1) \sqrt{1 + \dot{x}_1^2} +
\alpha(\varphi_2) \sqrt{1 + \dot{x}_2^2}],
\end{equation}
where $f$ is the bulk free energy density,
$\varphi_i = \mbox{atan}(\dot{x}_i)$ (with $\dot{x}_i = dx_i/dt$)
are the angles between the interfaces and the lattice $t$-axis, and
$\alpha(\varphi)$
is the angle dependent interface tension, which is known analytically
\cite{Rot81}. The minimal energy configuration contains two flat parallel
interfaces (with $\varphi_i = 0$) and it has
${\cal H}[x_1,x_2] = f L^2 + 2 \alpha(0) L$ independent of $M$.
The leading finite volume effect of $Z(M)$ is therefore given by Binder's
result
\begin{equation}
Z(M) \propto \exp(- \beta f L^2) \exp(- 2 \sigma L),
\label{partition}
\end{equation}
where $\sigma = \beta \alpha(0)$ is the reduced interface tension.
Eq.(\ref{partition}) receives corrections due to interface
fluctuations. Binder has parametrized this effect by a pre-exponential
factor $L^x$. One goal of this paper is to compute the value of $x$.
Since we are only interested in the next to leading finite volume effect
of $Z(M)$, we may assume that the interfaces are almost flat with only small
capillary wave fluctuations \cite{Buf65}, such that $\dot{x}_i^2 \ll 1$. Then
\begin{equation}
\beta {\cal H}[x_1,x_2] = \int_0^L dt [\beta f L + 2 \sigma +
\frac{\kappa}{2}\dot{x}_1^2 + \frac{\kappa}{2}\dot{x}_2^2],
\end{equation}
where $\kappa = \beta (\alpha(0) + \alpha''(0))$
is the reduced interface stiffness. (Note that $\alpha'(0) = 0$.)
The area $A = \int_0^L dt (x_1 - x_2)$ between the two interfaces determines
the value of the total magnetization $M = m (2A - L^2)$ of the configuration,
where $m$ is the magnetization density. For example, when the enclosed
area is maximal ($A = L^2$) the magnetization is $M = m L^2$, while for
$A = 0$ one has $M = - m L^2$.
For the partition function at fixed magnetization we now write the following
path integral
\begin{equation}
Z(M) = 2 \int {\cal D}x_1(t) {\cal D}x_2(t) \delta(\frac{M}{m} -
2 \int_0^L dt(x_1 - x_2) + L^2) \exp(- \beta {\cal H}[x_1,x_2]).
\label{pathintegral}
\end{equation}
The integration is over all pathes (interface locations) $x_1(t)$ and
$x_2(t)$, which are periodic in $t$ and which do not cross each other.
The non-crossing
constraint is a natural topological consequence, due to the fact that
the $x_i(t)$ are
interfaces separating the $+$ from the $-$ bulk phase. The factor 2 in front
arises
because the interfaces can wrap around the lattice in two different directions.
The system of eq.(\ref{pathintegral})
corresponds to two nonrelativistic particles (with coordinates $x_1(t)$ and
$x_2(t)$) on a circle of radius $L$, interacting repulsively at short
distances.
In addition the area enclosed between the pathes of the two particles is
constraint to $A = (M/m + L^2)/2$.
The path integral eq.(\ref{pathintegral})
is equivalent to a two-particle Schr\"odinger equation.
To solve the Schr\"odinger equation it is useful to
introduce a center of mass coordinate $X = (x_1 + x_2)/2$ and a relative
coordinate $x = x_1 - x_2$. The center of mass motion does not feel the
constraint. It is described by
plane waves with total momenta $P = 2 \pi l/L$ ($l \in \Z$) and with energies
$E_l = \beta f L + 2 \sigma + P^2/4 \kappa =$
$\beta f L + 2 \sigma + \pi^2 l^2/\kappa L^2$. The partition function
factorizes
\begin{equation}
Z(M) = 2 \exp(- \beta f L^2) \exp(- 2 \sigma L) \sum_{l \in \Z}
\exp(- \pi^2 l^2/\kappa L) z(M).
\end{equation}
Since we are interested in the large volume limit we may replace the sum by
an integral
$\sum_{l \in \Z} \exp(- \pi^2 l^2/\kappa L) \sim$ $\int_{-\infty}^\infty
dl \exp(- \pi^2 l^2/\kappa L) =$ $\sqrt{\kappa L/\pi}$.
The relative motion is described by
\begin{eqnarray}
z(M) & = & \int {\cal D}x(t) \delta(\frac{M}{m} - 2 \int_0^L dt \; x + L^2)
\exp(- \int_0^L dt \frac{\kappa}{4}\dot{x}^2) \nonumber \\
& = & \frac{1}{2 \pi} \int_{-\infty}^\infty d\theta \exp(i \theta \frac{M}{m})
\int {\cal D}x(t)
\exp(- \int_0^L dt [\frac{\kappa}{4}\dot{x}^2 +
2i\theta(x - \frac{L}{2})]).
\end{eqnarray}
In the last step the constraint of fixed magnetization has been rewritten as
an integral over an additional parameter $\theta$. The path integral for the
relative motion corresponds to a Schr\"odinger equation for a wave function
$\Psi_n(x)$ with an imaginary linear potential
\begin{equation}
- \frac{1}{\kappa} \frac{d^2 \Psi_n(x)}{dx^2} +
2i\theta(x - \frac{L}{2})\Psi_n(x) = \epsilon_n(\theta) \Psi_n(x).
\label{Schroedinger}
\end{equation}
The energies $\epsilon_n(\theta)$ are in general complex. The non-crossing
constraint of the interfaces reflects itself in the boundary conditions
$\Psi_n(0) = \Psi_n(L) = 0$. (One may also impose other boundary conditions
without changing the final result.) The partition function of the relative
motion is given by
\begin{equation}
z(M) = \frac{1}{2 \pi} \int_{-\infty}^\infty d\theta \exp(i\theta\frac{M}{m})
\sum_n \exp(- \epsilon_n(\theta) L),
\label{part}
\end{equation}
and the problem reduces to the computation of the $\theta$-dependent energy
spectrum. Since we are interested in small $|M|$, we need $\epsilon_n(\theta)$
only
for large values of $|\theta|$. The Schr\"odinger equation (\ref{Schroedinger})
is solved in terms of Airy functions $\Psi_n(x) = a \mbox{Ai}(z_n(x)) +
b \mbox{Bi}(z_n(x))$ with $z_n(x) = (\kappa/4 \theta^2)^{1/3}
(\epsilon_n(\theta) - 2i\theta(x - L/2))$. The boundary conditions result in
the
quantization condition
$\mbox{Ai}(z_n(0)) \mbox{Bi}(z_n(L)) =$ $\mbox{Ai}(z_n(L)) \mbox{Bi}(z_n(0))$.
For large values of $|\theta|$ the solutions come in pairs with complex
conjugate energies and one finds
\begin{equation}
\epsilon_n(\theta) \sim \pm i|\theta| L + (4 \theta^2/\kappa)^{1/3} r_n
\exp(\pm 2 \pi i/3),
\end{equation}
where the $r_n < 0$ are the zeros of the Airy function $\mbox{Ai}(r_n) = 0$.
In the large volume limit one can use the asymptotic form $r_n \sim
- (3\pi n/2)^{2/3}$ and one can replace the sum over $n$ in eq.(\ref{part})
by an integral. After some manipulations one finds
\begin{eqnarray}
z(M) &=& \frac{1}{\pi^2 \kappa L^3}
\int_0^\infty d\theta \cos(\theta \frac{M}{m})
\int_0^\infty dy \sqrt{y} \exp(- y/2 \kappa L) \times \nonumber \\
&&[\cos(\theta L^2) \cos(\sqrt{3}y/2 \kappa L) +
\sin(\theta L^2) \sin(\sqrt{3}y/2 \kappa L)]/\theta,
\end{eqnarray}
where $y = (3 \pi n \kappa \theta L^3)^{2/3}$. Using
$\int_0^\infty dy
\sqrt{y} \exp(- y/2 \kappa L) \cos(\sqrt{3}y/2 \kappa L) =$ $0$ and
$\int_0^\infty dy
\sqrt{y} \exp(- y/2 \kappa L) \sin(\sqrt{3}y/2 \kappa L) =$
$(\kappa L)^{3/2} \sqrt{\pi}/2$
one obtains
\begin{equation}
z(M) = \frac{1}{2 L^2} \sqrt{\frac{\kappa L}{\pi}} \frac{1}{\pi}
\int_0^\infty d\theta \cos(\theta \frac{M}{m}) \sin(\theta L^2)/\theta
= \frac{1}{4 L^2} \sqrt{\frac{\kappa L}{\pi}}
\end{equation}
for $|M| < m L^2$, and $z(M) = 0$ for $|M| > m L^2$.
Still, in the small $|M|$ region, the partition function is
independent of the magnetization. This is because the fluctuations of the
interfaces are small and do not induce interactions between the
two interfaces. In total one finds
\begin{equation}
Z(M) = \exp( - \beta f L^2) \frac{\kappa}{2 \pi L} \exp(- 2 \sigma L)
\;\;\;\;\; \mbox{for $|M| < m L^2$}.
\label{result}
\end{equation}
This corresponds to $x = -1$ in Binder's pre-exponential factor $L^x$.
Eq.(\ref{result}) is reliable only for small values of $|M|$,
because only in this
region the typical configurations consist of two domains separated by two
interfaces. When $|M|$ increases and one of the domains shrinks in favor of the
other one, the lowest energy configuration contains a droplet of one
phase in a background of the other phase. The equilibrium shape of the
droplet follows from the so-called Wulff construction \cite{Wul01}. Using the
analytic expression for the angle dependent interface tension
$\alpha(\varphi)$,
this construction has been carried out for the $d = 2$ Ising model in
ref.\cite{Rot81}. The equilibrium droplet has a spherical shape only close
to the second order phase transition, where a continuum limit is approached
and $\alpha(\varphi)$ becomes angle independent. In this case the region where
a droplet costs less energy than two parallel interfaces is at $|M| >
m L^2(1 - 2/\pi)$. In the zero temperature limit, on the other hand,
the equilibrium droplet becomes quadratic. Then a droplet is energetically
favorable for $|M| > m L^2/2$. The general case lies between the two
extremes. Of course, also a droplet has capillary wave fluctuations. They
could be handled in the same way as before. However, away from the critical
region the angle dependence of $\alpha(\varphi)$ complicates the situation.
In any case, we are most interested in the small $|M|$ region where the results
from above apply.

In the numerical simulations of the multimagnetical ensemble one measures
$Z(M)$ only up to a global normalization. In practice the reduced interface
tension $\sigma$ is extracted from the ratio of the partition function
at the minimum $Z(0)$ and at the maximum $Z(m L^2)$. At the maximum typical
configurations consist of just one domain (of either $+$ or $-$ phase) and no
interfaces are present. Such pure phase configurations have so far
not been included in the calculation. It is trivial to include
these configurations, because they only contribute a $\delta$-function to the
$M$-dependent
partition function. The $\delta$-function is normalized by the Boltzmann factor
of the free energy, and we can write
\begin{equation}
Z(M) = \exp(- \beta f L^2) [\delta(\frac{M}{m} - L^2) +
\delta(\frac{M}{m} + L^2) +
\frac{\kappa}{2 \pi L} \exp(- 2 \sigma L)].
\end{equation}
Of course, also a pure phase has fluctuations in the total magnetization,
which have not yet been taken into account.
It is well known that
the $\delta$-functions are smeared to Gaussian distributions
$D(M'/m) = \sqrt{1/2 \pi \chi L^2} \exp(- M'^2/2 m^2 \chi L^2)$
in a finite volume (see e.g. \cite{Bin81}).
Here $\chi$ is the magnetic susceptibility and $M' = M \pm m L^2$.
The same fluctuations should be included in the two phase configurations at
small $|M|$. This is achieved by the convolution
\begin{eqnarray}
Z(M)&\rightarrow&\int_{-\infty}^\infty d\frac{M'}{m}
Z(M - M') D(\frac{M'}{m}) \nonumber \\
& = & \exp(- \beta f L^2)
[D(\frac{M}{m} - L^2) + D(\frac{M}{m} + L^2) +
\frac{\kappa}{2 \pi L} \exp(- 2 \sigma L)].
\end{eqnarray}
This is the final result, which is valid in the two interface regime
at small $|M|$ as well as near the pure phase regions $M = \pm m L^2$. At
intermediate $|M|$, in the region where droplets dominate, it is not valid
and we will not use it there. Most interesting for applications to
numerical simulations is the ratio
\begin{equation}
\frac{Z(0)}{Z(m L^2)} = \sqrt{\frac{\chi}{2 \pi}} \kappa
\exp(- 2 \sigma L) \;\;\;\;\; \mbox{for $d = 2$},
\label{final}
\end{equation}
which has an $L$-independent pre-exponential factor.
The $L$-independence was already observed empirically in
ref.\cite{Ber91} by fitting the numerical data. Eq.(\ref{final}) allows to
determine the interface stiffness $\kappa$ from a fit of numerical data, when
$\chi$ is known.

Now let us turn to the $d = 3$ case. It is important to note that
3-dimensional interfaces living on a lattice have a roughening transition
(see e.g. \cite{Abr87}).
At very low temperatures the interfaces are rigid, with small steplike
excitations following the lattice structure. At temperatures above the
roughening transition, which is well below the bulk critical temperature,
interfaces become rough and fluctuate more freely.
\footnote{Note that in two
dimensions interfaces are always rough, except at zero temperature.}
In particular, the capillary
waves arise as soft modes of interface fluctuations. Here we restrict ourselves
to temperatures above the roughening transition. First, let us consider a
volume $L \times L \times L'$ with $L' \ll L$. Then capillary waves in the
short direction cost much energy and are therefore suppressed, such that
the problem
remains quasi 2-dimensional. Repeating the whole calculation with the extra
factor of $L'$ one obtains
\begin{equation}
Z(M) = \exp(- \beta f L^2 L')
[D(\frac{M}{m} - L^2 L') + D(\frac{M}{m} + L^2 L') +
\frac{\kappa}{2 \pi L} \exp(- 2 \sigma L L')],
\label{3d}
\end{equation}
where now
$D(M'/m) = \sqrt{1/2 \pi \chi L^2 L'}
\exp(- M'^2/2 m^2 \chi L^2 L')$.
Again this corresponds to $x = -1$ in Binder's pre-exponential factor.
When $L'$ increases and the problem becomes truely 3-dimensional,
eq.(\ref{3d}) is no longer correct, because capillary waves can also arise
in the third direction. As a consequence, numerical factors will change, but
the
general form of the $L$-dependence should remain the same. In the
$L' \rightarrow L$ limit
I therefore conjecture for the partition function ratio
\begin{equation}
\frac{Z(0)}{Z(m L^3)} \propto \sqrt{L}
\exp(- 2 \sigma L^2) \;\;\;\;\; \mbox{for $d = 3$},
\end{equation}
such that the pre-exponential factor is now $L$-dependent. This behavior
is in good
agreement with the numerical data of ref.\cite{Ber91} for large volumes.
The knowledge of
the pre-exponential behavior fixes one free parameter of the fit in
ref.\cite{Ber91}, and therefore improves the analysis of the numerical data.
In dimensions $d \geq 4$ it is generally believed that interfaces are
always rigid. Therefore one must be careful in generalizing the results to
higher dimensions. It is straightforward to generalize the results to other
models. Especially, the form of the $L$-dependence should be universal.

I would like to thank P. Hasenfratz for very interesting discussions.
I am also indepted to
V. Privman for explaining the concept of capillary waves, and for very useful
remarks.

\end{document}